\PassOptionsToPackage{unicode}{hyperref}
\PassOptionsToPackage{hyphens}{url}
\PassOptionsToPackage{dvipsnames,svgnames,x11names}{xcolor}
\documentclass[
  sn-vancouver,
  Numbered,
  ]{sn-jnl}

\usepackage{xcolor}
\usepackage{amsmath,amssymb}
\usepackage{iftex}
\ifPDFTeX
  \usepackage[T1]{fontenc}
  \usepackage[utf8]{inputenc}
  \usepackage{textcomp} 
\else 
  \usepackage{unicode-math} 
  \defaultfontfeatures{Scale=MatchLowercase}
  \defaultfontfeatures[\rmfamily]{Ligatures=TeX,Scale=1}
\fi
\usepackage{lmodern}
\ifPDFTeX\else
\fi
\IfFileExists{upquote.sty}{\usepackage{upquote}}{}
\IfFileExists{microtype.sty}{
  \usepackage[]{microtype}
  \UseMicrotypeSet[protrusion]{basicmath} 
}{}
\makeatletter
\@ifundefined{KOMAClassName}{
  \IfFileExists{parskip.sty}{%
    \usepackage{parskip}
  }{
    \setlength{\parindent}{0pt}
    \setlength{\parskip}{6pt plus 2pt minus 1pt}}
}{
  \KOMAoptions{parskip=half}}
\makeatother
\makeatletter
\ifx\paragraph\undefined\else
  \let\oldparagraph\paragraph
  \renewcommand{\paragraph}{
    \@ifstar
      \xxxParagraphStar
      \xxxParagraphNoStar
  }
  \newcommand{\xxxParagraphStar}[1]{\oldparagraph*{#1}\mbox{}}
  \newcommand{\xxxParagraphNoStar}[1]{\oldparagraph{#1}\mbox{}}
\fi
\ifx\subparagraph\undefined\else
  \let\oldsubparagraph\subparagraph
  \renewcommand{\subparagraph}{
    \@ifstar
      \xxxSubParagraphStar
      \xxxSubParagraphNoStar
  }
  \newcommand{\xxxSubParagraphStar}[1]{\oldsubparagraph*{#1}\mbox{}}
  \newcommand{\xxxSubParagraphNoStar}[1]{\oldsubparagraph{#1}\mbox{}}
\fi
\makeatother

\usepackage{longtable,booktabs,array}
\usepackage{calc} 
\usepackage{etoolbox}
\makeatletter
\patchcmd\longtable{\par}{\if@noskipsec\mbox{}\fi\par}{}{}
\makeatother
\IfFileExists{footnotehyper.sty}{\usepackage{footnotehyper}}{\usepackage{footnote}}
\makesavenoteenv{longtable}
\usepackage{graphicx}
\makeatletter
\newsavebox\pandoc@box
\newcommand*\pandocbounded[1]{
  \sbox\pandoc@box{#1}%
  \Gscale@div\@tempa{\textheight}{\dimexpr\ht\pandoc@box+\dp\pandoc@box\relax}%
  \Gscale@div\@tempb{\linewidth}{\wd\pandoc@box}%
  \ifdim\@tempb\p@<\@tempa\p@\let\@tempa\@tempb\fi
  \ifdim\@tempa\p@<\p@\scalebox{\@tempa}{\usebox\pandoc@box}%
  \else\usebox{\pandoc@box}%
  \fi%
}
\def\fps@figure{htbp}
\makeatother

\providecommand{\tightlist}{%
  \setlength{\itemsep}{0pt}\setlength{\parskip}{0pt}}

\usepackage{graphicx}%
\usepackage{multirow}%
\usepackage{amsmath,amssymb,amsfonts}%
\usepackage{amsthm}%
\usepackage{mathrsfs}%
\usepackage[title]{appendix}%
\usepackage{xcolor}%
\usepackage{textcomp}%
\usepackage{manyfoot}%
\usepackage{booktabs}%
\usepackage{algorithm}%
\usepackage{algorithmicx}%
\usepackage{algpseudocode}%
\usepackage{listings}%

\usepackage{booktabs}
\usepackage{longtable}
\usepackage{array}
\usepackage{multirow}
\usepackage{wrapfig}
\usepackage{float}
\usepackage{colortbl}
\usepackage{pdflscape}
\usepackage{tabu}
\usepackage{threeparttable}
\usepackage{threeparttablex}
\usepackage[normalem]{ulem}
\usepackage{makecell}
\usepackage{xcolor}
\usepackage{dsfont}
\makeatletter
\@ifpackageloaded{caption}{}{\usepackage{caption}}
\AtBeginDocument{%
\ifdefined\contentsname
  \renewcommand*\contentsname{Table of contents}
\else
  \newcommand\contentsname{Table of contents}
\fi
\ifdefined\listfigurename
  \renewcommand*\listfigurename{List of Figures}
\else
  \newcommand\listfigurename{List of Figures}
\fi
\ifdefined\listtablename
  \renewcommand*\listtablename{List of Tables}
\else
  \newcommand\listtablename{List of Tables}
\fi
\ifdefined\figurename
  \renewcommand*\figurename{Figure}
\else
  \newcommand\figurename{Figure}
\fi
\ifdefined\tablename
  \renewcommand*\tablename{Table}
\else
  \newcommand\tablename{Table}
\fi
}
\@ifpackageloaded{float}{}{\usepackage{float}}
\floatstyle{ruled}
\@ifundefined{c@chapter}{\newfloat{codelisting}{h}{lop}}{\newfloat{codelisting}{h}{lop}[chapter]}
\floatname{codelisting}{Listing}

\makeatother
\makeatletter
\@ifpackageloaded{caption}{}{\usepackage{caption}}
\@ifpackageloaded{subcaption}{}{\usepackage{subcaption}}
\makeatother
\usepackage{bookmark}
\IfFileExists{xurl.sty}{\usepackage{xurl}}{} 
\hypersetup{
  pdftitle={Initial data analysis of the national German transplantation registry with a focus on kidney transplantation},
  pdfauthor={Lukas Klein; Gunter Grieser; Carl-Ludwig Fischer-Fröhlich; Axel Rahmel; Henrik Stahl; Andreas Wienke; Antje Jahn-Eimermacher},
  pdfkeywords={Kidney transplantation, German transplantation
registry, Initial data analysis, Missing data, Survival analysis},
  colorlinks=true,
  linkcolor={blue},
  filecolor={Maroon},
  citecolor={Blue},
  urlcolor={Blue},
  pdfcreator={LaTeX via pandoc}}

\title[Initial data analysis of the national German transplantation
registry with a focus on kidney transplantation]{Initial data analysis
of the national German transplantation registry with a focus on kidney
transplantation}

\author*[1,2]{\fnm{Lukas} \sur{Klein}}\email{lukas.klein@h-da.de}\author[2]{\fnm{Gunter} \sur{Grieser}}\author[3]{\fnm{Carl-Ludwig} \sur{Fischer-Fröhlich}}\author[3]{\fnm{Axel} \sur{Rahmel}}\author[2]{\fnm{Henrik} \sur{Stahl}}\author[1]{\fnm{Andreas} \sur{Wienke}}\author[2]{\fnm{Antje} \sur{Jahn-Eimermacher}}
\affil[1]{, \orgname{Martin Luther University
Halle-Wittenberg}, \orgaddress{\street{Magdeburger Straße
8}, \city{Halle}, \postcode{6112}}}
\affil[2]{, \orgname{Darmstadt University of Applied
Sciences}, \orgaddress{\street{Schöfferstraße
3}, \city{Darmstadt}, \postcode{64295}}}
\affil[3]{, \orgname{Deutsche Stiftung
Organtransplantation}, \orgaddress{\street{Deutschherrnufer
52}, \city{Frankfurt am Main}, \postcode{60594}}}

\abstract{This study presents an Initial Data Analysis (IDA) of the
German Transplantation Registry (TxReg) data for a better data
understanding and to inform future data analyses. The IDA is focusing on
data on first-time kidney-only transplantations in adult recipients from
deceased donors between 2006 and 2016 and refers to data from 14,954
recipients and 9,964 donors across 25 tables. Investigated aspects
include missing data patterns and structure, data consistency, and
availability of event time data. Results show that missing data
proportions vary widely, with some tables nearly complete while others
have over 50\% missing values. Missing data patterns are identified
using a decision tree approach. An influx and outflux analysis
demonstrates that some variables have high potential for imputing
missing data, while others were less suitable for imputation. We
identified 168 multi-sourced variables that are reported by multiple
data providers in parallel leading to discrepancies for some variables
but also providing opportunities for missing data imputation. Our
findings on event time data demonstrate the importance of carefully
selecting the variables used for event time analyses as results will
strongly depend on this selection. In summary, our findings highlight
the challenges when utilizing the TxReg data for research and provide
recommendations for data preprocessing and analysis in future analyses.}


\begin{document}
\maketitle

\section{Background}\label{background}

Since 2021, the \emph{German Transplantation Registry (TxReg)} is the
central archive for data on all solid-organ transplantations in Germany.
It was established in the German transplantation law
(Transplantationsgesetz, TPG) in 2016 \citep{TPG} to collect data from
the three data providers which coordinate the transplantation processes
in Germany. The \emph{Eurotransplant Foundation (ET)} is responsible for
the allocation of organs in Germany, Austria, Belgium, Croatia, Hungary,
Luxembourg, the Netherlands and Slovenia. In Germany, the \emph{German
Organ Transplantation Foundation (DSO)} is responsible for the
coordination of organ donation and the \emph{Institute for Quality
Assurance and Transparency in Health Care (IQTIG)} is responsible for
the quality assurance of the transplantation process. By national law,
transplantation centers must report data of every transplantation
procedure for the first three years of follow-up annually to IQTIG. For
allocation, recipient data, as well as long-term follow-up data -
especially in the context of retransplantation - must be stored at ET
and transplantation centers. Therefore, transplantation data flows
parallel through all three data providers to the TxReg.

The main goal of the TxReg is to improve the transparency and quality of
the transplantation process by providing data for research and quality
assurance \citep{txreg}. However, the data has not been widely used for
research purposes yet. At the time of writing, the official website of
the registry only shows nine publications using the data. A potential
reason for that might be the complexity of the data structure caused by
the collection process from multiple data providers and potential
quality issues of the data \citep{Otto2024}.

Before data is used to answer research questions, it is important to
conduct an \emph{initial data analysis (IDA)}. The IDA is used to assess
the quality of the data, to identify potential problems, and to
determine the data's ability to address the planned research objectives.
The STRATOS initiative provides guidelines for an IDA. In this paper, we
follow these recommendations, with an emphasis on transparency and
reproducibility, clear visualizations and summaries of the data, a
systematic assessment of missing data, and early consideration of IDA
findings to inform subsequent analyses \citep{Baillie2022}.

As a basis for future research, this paper provides an overview of the
data structure of the TxReg data to enable future studies to utilize it
more effectively. We first analyze missing data patterns and structure
using flux analysis and a decision tree approach. Flux analysis provides
insight into the potential for missing data imputation, while decision
trees predict missingness patterns based on other variables to help
identify potential missing data mechanisms. We then analyze the
consistency of information between variables contributed by different
data providers and assess imputation potential. Finally, we provide
guidance for using data in event time analysis of long-term outcomes, as
e.g.~graft and patient survival.

We focus on the kidney data from the TxReg. Kidney failure, also known
as end-stage kidney disease (ESKD), is a severe condition describing a
complete loss of kidney function. It develops from chronic kidney
disease (CKD) as the terminal stage. Common causes for CKD and ESKD
include hypertension, other cardiovascular diseases, glomerular
diseases, diabetes mellitus, cystic kidney diseases, urinary tract
diseases, and congenital birth defects \citep{Hashmi2023}. Diagnosis of
ESKD is based on the glomerular filtration rate (GFR)
\citep{Hashmi2023}.

Long-term treatment options for ESKD include dialysis and kidney
transplantation \citep{Bello2024}. Compared to dialysis, kidney
transplant recipients show a higher quality of life and a lower
mortality rate. These benefits are especially pronounced when a
transplantation is performed before starting dialysis
\citep{Amaral2016}. The availability of kidney transplantation as a
treatment, however, is greatly limited by the number of available
organs. In Germany in 2023, 2,617 registrations to the waiting list for
a kidney transplantation were made \citep{DSO2023}. In the same year,
2,122 kidney transplantations were performed. However, at the end of
2023, 10,454 patients were still on the waiting list \citep{DSO2023}.
Before a kidney transplantation can be performed, compatibility between
the donor and recipient needs to be assured to prevent graft rejection
and other complications. To optimize treatment effects both on a
individual and a population level, an allocation algorithm is used to
find recipients for a donor organ. Data from the TxReg allow
e.g.~important insight into the consequences of such organ allocation
rules for certain patient groups \citep{Kolbrink2024}.

When analyzing outcomes of kidney transplantation, there are different
endpoints, which can be derived. When looking at graft failure, the
following endpoints are common \citep{EBPG2002}:

\begin{enumerate}
\def\labelenumi{\arabic{enumi}.}
\tightlist
\item
  \textbf{Patient survival}: Time from transplantation to death;
  individuals alive at last follow-up are censored at that time.
\item
  \textbf{Graft and patient survival}: Time from transplantation to
  irreversible graft failure (return to long-term dialysis or
  retransplantation) or death.
\item
  \textbf{Death-censored graft survival}: Time from transplantation to
  irreversible graft failure, with death while the graft is still
  functioning treated as a censoring event at death.
\end{enumerate}

We start with notations and definitions including a description of the
data structure of the TxReg data. We then describe the methods used for
an analysis of missing data, data consistency and distribution of event
time data. After presenting the results, we end with a discussion and
conclusion.

\newcommand{\tname}{\mathbf{X}}
\newcommand{\inst}{\text{DP}}
\newcommand{\instj}{{\inst}_{\tname j}}
\newcommand{\instk}{{\inst}_{\tname k}}
\newcommand{\colset}{\mathcal{C}_\tname(\inst)}
\newcommand{\obset}{\mathcal{I}_\tname(\inst)}
\newcommand{\ET}{\text{ET}}
\newcommand{\DSO}{\text{DSO}}
\newcommand{\IQTIG}{\text{IQTIG}}
\newcommand{\textmis}{\textit{ is missing}}
\newcommand{\mis}{\tname_{i,j} \textmis}
\newcommand{\textnmis}{\textit{ is not missing}}
\newcommand{\nmis}{\tname_{i,j} \textnmis}
\newcommand{\obsetj}{\mathcal{I}_\tname(\instj)}
\newcommand{\ninst}{|\obset|}
\newcommand{\colsetj}{\mathcal{C}_\tname(\instj)}
\newcommand{\ind}[1]{\mathds{1}\left(#1\right)}

\section{Methods}\label{sec-methods}

\subsection{Notations and Definitions}\label{sec-notations}

The organization that is managing the TxReg data \citep{techspec}
internally uses four relational databases to collect, store, track and
export data. For the export (to which we refer as the TxReg data), the
data is provided in the form of a relational database, exported as
comma-separated values (CSV) files. In the official documentation of the
TxReg, three entities are described: transplantations, recipients and
donors. Each of these entities in the database has a unique identifier
provided by ET. In the Supplementary Table 1, Additional File 1, we
provide an overview of the tables in the TxReg data and to which subject
types they refer. Linkage between those entities is possible via the
pseudonymized identifiers provided by ET \citep{txregHandbuch2025}.

When the data is provided to third parties for research purposes, it is
anonymized. That includes encryption of identifiers and removal of
personal information. Since certain dates could allow re-identification
of patients, dates are provided as relative dates w.r.t. some
undisclosed reference date. This allows to derive the time interval
between transplantation and graft failure e.g.~without exposing exact
dates.

Each table contains columns and observations (rows). For the ease of
reading we denote tables and columns by a prefix \texttt{T\_} and
\texttt{C\_}, respectively. For example,
\texttt{T\_Transplantation.C\_Destination} refers to the column
\texttt{Destination} in the table \texttt{Transplantation}. To allow
international readability we use translated names for the columns, which
are provided in the Supplementary Table 2, Additional File 1.

For a table \(\mathbf{X}\) we denote the number of observations and
columns by \(n_{\mathbf{X}}\) and \(p_{\mathbf{X}}\), respectively.

To avoid ambiguity with transplantation centers, we refer to
Eurotransplant (ET), the German Organ Transplantation Foundation (DSO),
and the Institute for Quality Assurance and Transparency in Health Care
(IQTIG) collectively as data providers (DPs). In formulas, the symbol
\(\text{DP}\) denotes the data provider contributing a column or
observation.

All tables can contain missing values. However, especially for
qualitative test results, explicit values like ``not tested'' or
``unknown'' exist, which we do not consider as missing values in the
context of an IDA. We acknowledge that, depending on the variable and
clinical context, such categories may in some settings be treated as
missing values. In this analysis, however, we consistently treat them as
informative categories rather than missing data.

Three data patterns (namely multi-sourced variables, relating
observations and data providers and the mixing of cross-sectional and
longitudinal data) play an important role for our IDA, we describe it
them in the next subsections.

\subsubsection{Multi-Sourced Variables}\label{sec-multivars}

Each column \(\mathbf{X}_j\), \(j=1,...,p_{\mathbf{X}}\), of table
\(\mathbf{X}\) is contributed by one of the data providers: ET, DSO or
IQTIG. Supplementary Table 5, Additional File 1 shows to which tables
which data provider contributed data. Thereby, not all columns may
represent unique variables, due to the data collection process through
three data providers. For example, the donor's weight is stored in three
columns, one from each data provider. We refer to such variables as
``multi-sourced variables''. Their columns share the same name as a
prefix and give the data provider identifier as a suffix. Multi-sourced
variables can yield conflicting values but can also provide a chance to
impute missing values.

To denote which data provider provides a column we use the following
notation: Let \({\text{DP}}_{\mathbf{X}j}\) be the data provider that
contributes the \(j\text{th}\) column of the table \(\mathbf{X}\). Let
then \(\mathcal{C}_\mathbf{X}(\text{DP})\) be the set of all indices of
columns contributed by data provider
\(\text{DP}\in \{\text{ET},\text{DSO}, \text{IQTIG}\}\) to table
\(\mathbf{X}\). This set is defined as follows:

\[
\mathcal{C}_\mathbf{X}(\text{DP})= \{ j \in \{1, \dots, p_{\mathbf{X}}\} \;|\; {\text{DP}}_{\mathbf{X}j}= \text{DP}\}.
\]

For example, in \texttt{T\_Recipient},
\(\mathcal{C}_\mathbf{X}(\text{ET})\) contains all columns with suffix
\texttt{ET} (e.g., \texttt{EBasisGewichtWertET},
\texttt{EBasisGroesseWertET}), while
\(\mathcal{C}_\mathbf{X}(\text{IQTIG})\) contains the corresponding
IQTIG-reported columns (e.g., \texttt{EBasisGewichtWertIQTIG},
\texttt{EBasisGroesseWertIQTIG}). A complete list of all multi-sourced
variable pairs is provided in the digital supplementary material.

\subsubsection{Relating Observations to Data
Providers}\label{sec-uncobs}

Not only variables, but also observations are contributed by all three
data providers.

A data provider contributes an observation to the table if at least one
of the columns contributed by the data provider has a non-missing value.
Therefore we define \(\mathcal{I}_\mathbf{X}(\text{DP})\) to be the set
of all indices of observations in table \(\mathbf{X}\) contributed by
data provider \(\text{DP}\) as follows:

\[
\mathcal{I}_\mathbf{X}(\text{DP})= \left\{
 i \in \{1, \dots, n_{\mathbf{X}}\} \;|\;
  \exists j \in \mathcal{C}_\mathbf{X}(\text{DP})\text{ such that } \mathbf{X}_{i,j} \textit{ is not missing}
\right\}.
\]

In practical terms, an observation can belong to more than one
observation set \(\mathcal{I}_\mathbf{X}(\text{DP})\) (for different
values of \(\text{DP}\)), depending on which provider-specific columns
are non-missing in that observation. For example, in
\texttt{T\_Transplantation}, an observation with non-missing values in
both ET and IQTIG columns would belong to both
\(\mathcal{I}_\mathbf{X}(\text{ET})\) and
\(\mathcal{I}_\mathbf{X}(\text{IQTIG})\). Therefore we consider
provider-specific ``sub-tables'' within each TxReg table: for a given
table, we analyze ET, DSO, and IQTIG contributions separately. Columns
will always belong to a specific provider through
\(\mathcal{C}_\mathbf{X}(\text{DP})\), while observations can belong to
multiple providers through \(\mathcal{I}_\mathbf{X}(\text{DP})\). This
directly reflects the TxReg data structure and avoids counting
structurally absent entries as missing values.

\subsubsection{Cross-Sectional and Longitudinal Data}\label{sec-mix}

Some tables contain cross-sectional data, where each observation
represents a single data entry, independent of the time of data
collection. Other tables contain longitudinal data, where each
observation represents a single data entry at a specified time point.
The longitudinal data usually are complemented by a time identifier.
Some tables contain both cross-sectional and longitudinal data. For
example for the table \texttt{T\_Transplantation}, ET contributes
cross-sectional data, while IQTIG contributes longitudinal data. While
ET uses an ET identifier for the identification of the transplantation
entity, IQTIG uses in some cases the ET recipient identifier, in some
cases the ET donor identifier and in some cases both identifiers. This
leads to a situation where within the same table, multiple different
classes of entities are present. Important to note here is that IQTIG
does not report a time point for most observations. In these tables,
column based data analysis for the IDA can lead to misleading results,
because the number of observations is no longer meaningful in these
tables. For example, the table \texttt{T\_Transplantation} contains many
more observations than transplantations as some transplantations are
reported in more than one row. The correct number of transplantations
however can be derived from the number of observations contributed by ET
(\(|\mathcal{I}_\mathbf{X}(\text{ET}|\)) with \(\mathbf{X}=
\text{T\_Transplantation}\). The number of observations for each entity
is shown in Supplementary Table 6 to 8, Additional File 1. For this
reason in our IDA we analyze the columns contributed by ET, DSO and
IQTIG separately for each table.

\subsection{Target Population}\label{target-population}

We are using the ``legacy data'' (Altdaten), comprising data collected
between 2006 and 2016, i.e., before the TPG amendment 2016, which
introduced the requirement for informed consent for data collection for
research purposes.

The TxReg data primarily cover donors, transplantations, and recipients
in Germany. Within Eurotransplant, however, a small number of postmortem
donated organs from Germany are transplanted abroad, and conversely, a
small number of organs donated abroad are transplanted in Germany.
Geographically, the TxReg data therefore include donors from Germany,
transplantations in Germany, and recipients in Germany, with minor
cross-border exchanges.\citep{txregHandbuch2025}

The initial data analysis is focusing on kidney transplantations where
1. the recipient is adult, 2. it is the first transplantation for this
recipient and 3. the organ is from a deceased donor. The target
population was constructed using the ET-reported recipients, donors, and
transplantations in \texttt{T\_Recipient}, \texttt{T\_Transplantation},
\texttt{T\_Kidney\_Waitlist}, and \texttt{T\_Donor\_PM}. Recipients with
multiple transplantations were identified and removed using
\texttt{T\_Transplantation} itself, but also using
\texttt{T\_Kidney\_Waitlist.C\_KidneyTXCount}. We only considered
columns that had any data on this target population and removed
duplicated columns, which led to the final set of 1123 columns used in
the analyses. More details on the filtering process can be found in the
supplementary material under Supplementary Information 1 and
Supplementary Figure 1, Additional File 1. Our population consisted of
14,954 recipients and 9,964 donors across 25 tables.

\subsection{Missing Data analysis}\label{sec-miss}

Missing values need to be accounted for in analysis and modelling. Many
statistical and machine learning algorithms cannot handle missing values
and require imputation, data deletion or encoding of missing values.
Imputation often relies on the values of other columns, which are then
used as predictors \citep{Emmanuel2021}. The decision on how to handle
missing values is influenced by the number of missing values and the
underlying structure. The following methods are applied to guide such
decisions.

First, we analyze the proportion of missing values. \(\mathds{1}\) is
the indicator function. By \(M(\mathbf{X}_j)\) and \(R(\mathbf{X}_j)\)
we denote the number of missing and non-missing values for the
\(j\text{th}\) column of table \(\mathbf{X}\) contributed by data
provider \(\text{DP}\), respectively:

\[
M(\mathbf{X}_j)=\sum_{i \in \mathcal{I}_\mathbf{X}({\text{DP}}_{\mathbf{X}j})}{\mathds{1}\left(\mathbf{X}_{i,j} \textit{ is missing}\right)}
\]

\[
R(\mathbf{X}_j)=\sum_{i \in \mathcal{I}_\mathbf{X}({\text{DP}}_{\mathbf{X}j})}{\mathds{1}\left(\mathbf{X}_{i,j} \textit{ is not missing}\right)}
\]

Usually, the proportion of missing values is computed as the number of
missing values divided by the total number of observations
\(n_{\mathbf{X}}\). In our setting, this denominator is inappropriate,
because aspects of the data representation induce structural missingness
that does not reflect data quality. In addition to the mixed
cross-sectional/longitudinal structure described, further representation
issues occur: information from different data providers for the same
event is usually stored in a single row with data-provider-specific
columns, but in some instances it is stored across multiple rows.
Consequently, some entries are systematically missing by design
(representation/storage) rather than due to incomplete reporting.

Thus, instead of \(n_{\mathbf{X}}\) we use the number of observations
contributed by data provider \(\text{DP}\)
(\(|\mathcal{I}_\mathbf{X}(\text{DP})|\)). The proportion of missing
values \(PM(\mathbf{X}_j)\) for column \(j\text{th}\) in table
\(\mathbf{X}\) can then be calculated as:

\[
PM(\mathbf{X}_j)=\frac{M(\mathbf{X}_j)
}{ |\mathcal{I}_\mathbf{X}({\text{DP}}_{\mathbf{X}j})| }.
\]

For ET-contributed columns in a table, the denominator is the number of
observations with at least one non-missing ET value (not all rows of
that table) In a complex dataset, not only the amount of missing data,
but also the structure of the missing data is relevant for analysis. It
can reveal mechanisms and potential sources for biases. Two methods are
used to analyze the structures in the missing data. First, influx and
outflux analysis according to \citet{vanBuuren2018} is performed to
identify patterns of missing data. Second, to identify potential causal
mechanisms for missing data, a decision tree analysis to understand the
structure of the missing data is conducted based on the work by
\citet{Tierney2015}. The first approach works vertically on the columns,
while the second one operates horizontally on the observations.

\subsubsection{Influx and Outflux Analysis}\label{sec-flux}

The influx is a measure describing the potential for imputing missing
values in a column from not-missing values in other columns.
Accordingly, the outflux describes the potential of not-missing values
in a column to impute missing values in other columns.

When comparing two columns \(\mathbf{X}_{j}\) and \(\mathbf{X}_{k}\) of
table \(\mathbf{X}\) from the same data provider
\({\text{DP}}_{\mathbf{X}j}={\text{DP}}_{\mathbf{X}k}\), we can
calculate different statistics based on pairwise comparison of the
missingness within an observation. The proportion of usable cases
\(U(\mathbf{X}_j,\mathbf{X}_k)\) \citep{vanBuuren2018} between the
\(j\)th and \(k\)th column in \(\mathbf{X}\) from the same data provider
is the proportion of missing values of \(\mathbf{X}_{j}\) that could be
imputed by the use of non-missing observations in \(\mathbf{X}_{k}\). It
is defined as follows:

\[
U(\mathbf{X}_j,\mathbf{X}_k) = \frac{\sum_{i \in \mathcal{I}_\mathbf{X}({\text{DP}}_{\mathbf{X}j})}{\mathds{1}\left(\mathbf{X}_{i,j} \textit{ is missing}\wedge 
\mathbf{X}_{i,k}\textit{ is not missing}\right)}}{M(\mathbf{X}_j)} \quad \text{for } {\text{DP}}_{\mathbf{X}j}= {\text{DP}}_{\mathbf{X}k}
.
\]

This statistic can be used to check whether a column \(\mathbf{X}_{k}\)
has the potential to be used for missing data imputation of column
\(\mathbf{X}_{j}\). Influx \(I(\mathbf{X}_{j})\) and outflux
\(O(\mathbf{X}_{j})\) of column \(\mathbf{X}_j\) are the application of
the same approach to all columns from the same data provider of table
\(\mathbf{X}\).

The influx is the proportion of missing value in any column
\(k \in \mathcal{C}_\mathbf{X}({\text{DP}}_{\mathbf{X}j})\) that could
be imputed by the use of a non-missing observation in
\(\mathbf{X}_{j}\). The outflux is the proportion of non-missing value
in any column \(k \in\mathcal{C}_\mathbf{X}({\text{DP}}_{\mathbf{X}j})\)
that could be used to impute any missing value in \(\mathbf{X}_{j}\)
\citep{vanBuuren2018}.

\[
I(\mathbf{X}_{j})= \frac{\sum_{k \in \mathcal{C}_\mathbf{X}({\text{DP}}_{\mathbf{X}j})}\sum_{i \in \mathcal{I}_\mathbf{X}({\text{DP}}_{\mathbf{X}j})}\mathds{1}\left(\mathbf{X}_{i,j} \textit{ is missing}\wedge 
\mathbf{X}_{i,k} \textit{ is not missing}\right)}{\sum_{k \in \mathcal{C}_\mathbf{X}({\text{DP}}_{\mathbf{X}j})}R(\mathbf{X}_k)}
\]

\[
O(\mathbf{X}_{j})= \frac{\sum_{k \in \mathcal{C}_\mathbf{X}({\text{DP}}_{\mathbf{X}j})}\sum_{i \in \mathcal{I}_\mathbf{X}({\text{DP}}_{\mathbf{X}j})}\mathds{1}\left(\mathbf{X}_{i,j} \textit{ is not missing}\wedge 
\mathbf{X}_{i,k} \textit{ is missing}\right)}{\sum_{k \in \mathcal{C}_\mathbf{X}({\text{DP}}_{\mathbf{X}j})}M(\mathbf{X}_k)}
\]

These values depend on the number of missing values, therefore
comparisons of influx and outflux are usually not reasonable. We
visualize influx and outflux in a scatterplot. Only for
\texttt{T\_Recip\_Virology} it was not possible to calculate the
outflux, as there are no missing values in any column.

\subsubsection{Missing Data Structure Analysis}\label{sec-tree}

\citet{Tierney2015} proposed a method to better understand the structure
of missing data in that they investigate which columns in a table best
predict the number of missing values per observation. For this, they
proposed a regression tree. By analyzing the structure of the tree, it
can be identified which columns are important predictors for the number
of missing values per observation. This can help to understand the
mechanisms of missing data and to identify potential sources of bias or
improve downstream analyses.

The target variable for training the trees is the proportion of missing
values per observation. We calculate trees separately by data provider
for the reasons described. Thus, for a data provider \(\text{DP}\) the
target variable for observation
\(i \in \mathcal{I}_\mathbf{X}(\text{DP})\) is calculated as:

\[
OPM(\mathbf{X},\text{DP},i) = \frac{\sum_{j \in \mathcal{C}_\mathbf{X}(\text{DP})}\mathds{1}\left(\mathbf{X}_{i,j} \textit{ is missing}\right)}{|\mathcal{C}_\mathbf{X}(\text{DP})|} 
\]

Each table was split into a training and a test set with a 70/30 split.
The regression tree was trained on the training set and the accuracy of
the prediction was evaluated on the test set using the root mean squared
error (RMSE). The \texttt{rpart} \citep{Therneau1999} package was used
for tree building, as it supports surrogate splits. Training used
default parameters for variance splitting: a minimum of 20 observations
for a split, a minimum of 7 observations in a leaf, a complexity
parameter of 0.01, 5 maximum surrogates, majority vote if a surrogate
split is not possible, 10 cross-validation folds, and a maximum depth of
30. For training, we only considered columns with at least two distinct
values. Additionally, non-numeric columns were considered only if they
have at most 250 distinct values.

Adjusted feature importance was calculated to order the columns by their
importance for predicting the proportion of missing values per
observation. The importance of a predictor in \texttt{rpart} is
calculated by the sum of the goodness of split measures plus the
adjusted goodness of split measures for surrogate splits
\citep{Therneau1999}.

\subsection{Data Consistency}\label{data-consistency}

Beyond data completeness, data consistency represents a critical aspect
for reliable analyses. In particular, the multi-sourced variables,
present both analytical challenges and opportunities. While they cause
collinearity that negatively affects analyses e.g.~by increasing
standard errors, imputation strategies could highly benefit from using
this information: Missing values in one data provider's column can
potentially be imputed using values from another data provider's
corresponding column, thereby improving overall data completeness. We
therefore first examine the potential of multi-sourced variables for
imputation and second assess the degree of agreement between information
reported by several data providers. These analyses can inform both
decisions on data preprocessing and variable selection.

The proportion of usable cases \(U(\mathbf{X}_j,\mathbf{X}_k)\) will be
used to identify the potential of missing data imputation by the
multi-sourced variables. To analyze the consistency and agreement of
information between columns, Cramér's V and the Pearson correlation
coefficient are calculated. Cramér's V is a measure of association
between two categorical variables and is calculated from the chi-squared
statistic. It ranges from 0 (no association) to 1 (perfect association).
For two numerical variables, the absolute value of the Pearson
correlation is calculated, which ranges from 0 (no linear association)
to 1 (perfect linear association). The absolute value is used as we are
interested in the strength of the association.

\subsection{Distribution of Event Time
Data}\label{distribution-of-event-time-data}

A major potential of registries like the TxReg is the ability to study
long-term outcomes of solid organ transplantations, such as
time-to-event outcomes (e.g., patient death, graft failure, or
combinations of these) using survival analysis. We explore important
information given in TxReg to enable such analyses. Thereby, we focus on
event time analysis for the time from organ transplantation to patient
death and/or organ failure, that is a common outcome used in research on
transplantation strategies
\citep{liv_Brustia2020, kid_Sun2024, heart_Aleksova2020}.

For event time analyses of graft and patient survival, the following
information is needed: The date of organ transplantation, the date of
patient death, the date of organ failure and the date of last contact.
The latter is required to properly consider the time patients are at
risk and observed for experiencing events in estimates of incidence
rates and proportions. As already mentioned dates cannot be interpreted
as calendar dates, instead dates will refer to relative dates only.

These four dates needed for event time analyses are provided by
different data providers and are distributed across tables and columns
making the extraction of these dates ambiguous. We will explore the use
of different variables to derive these dates by visualizing their
distributions with kernel density estimates and illustrating
associations between variables by scatterplots.

All analyses and processing steps were implemented with Python,
Snakemake \citep{Mlder2021} and the \texttt{pandas}
\citep{reback2020pandas, mckinney-proc-scipy-2010} package.
Visualizations and further downstream analyses of results were
implemented with R and the \texttt{tidyverse} \citep{tidyverse} family
of packages. A full list of packages and versions used is available in
the Digital Supplementary material.

\section{Results}\label{sec-results}

\subsection{Missing Data Analysis}\label{sec-results-miss}

\begin{figure}[H]

\centering{

\includegraphics[width=3.5in,height=3.5in]{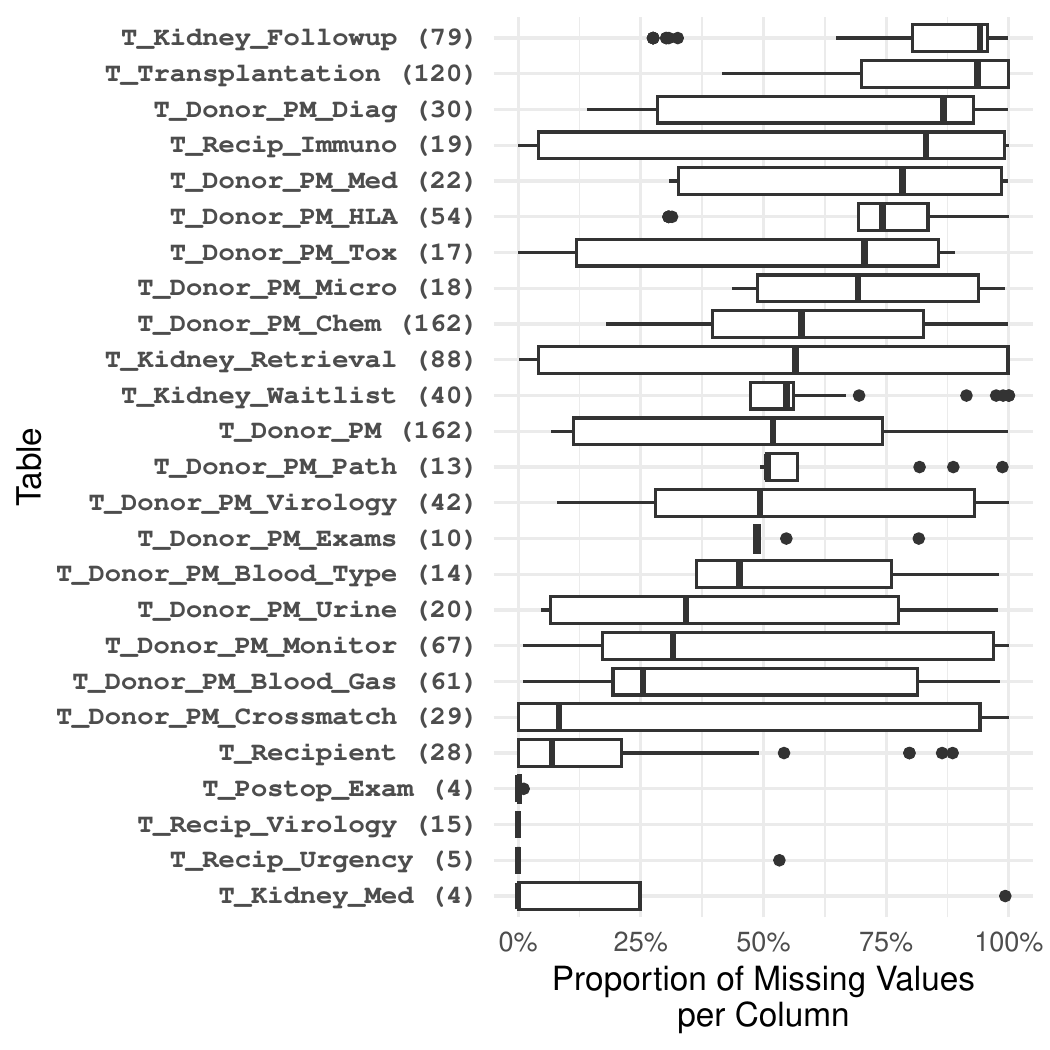}

}

\caption[Boxplot of the proportion of missing values per column for all
tables]{\label{fig-inst-missing}Boxplot of the proportion of missing
values in each column across all tables.}

\end{figure}

Figure~\ref{fig-inst-missing} illustrates the proportion of missing data
in each column of the tables \(PM(\mathbf{X}_j)\).

Our analysis reveals that \texttt{T\_Kidney\_Med},
\texttt{T\_Recip\_Urgency}, \texttt{T\_Recip\_Virology} and
\texttt{T\_Postop\_Exam} are mostly complete, with only a few columns
having missing values and their median proportion of missing values
being close to 0\%. For \texttt{T\_Recip\_Virology} this can be
explained with the use of explicit missing value indicators like ``not
tested''. The other tables have a small number of nearly complete
columns. Most of them are identifiers or time point information for
longitudinal data.

Our analysis also identifies that the tables with columns with high
proportions of missing values are \texttt{T\_Kidney\_Followup},
\texttt{T\_Transplantation}, \texttt{T\_Donor\_PM\_Diag},
\texttt{T\_Recip\_Immuno} and \texttt{T\_Donor\_PM\_Med}. When looking
at the data, all of these have in common that they contain longitudinal
data. Within these tables, each observation refers to a specific event
or lab test. For example, \texttt{T\_Kidney\_Followup} contains
follow-up events, which can be either initial or recurrent follow-ups.
\texttt{T\_Recip\_Immuno} contains lab tests for immunological
parameters, such as HLA typing or antibody screening. Depending on which
test or event is reported, different columns are used to report the
data, implying that other columns are missing. This could explain the
high proportions of missing observations per column in these tables.

\begin{figure}[H]

\centering{

\includegraphics[width=3.5in,height=3.5in]{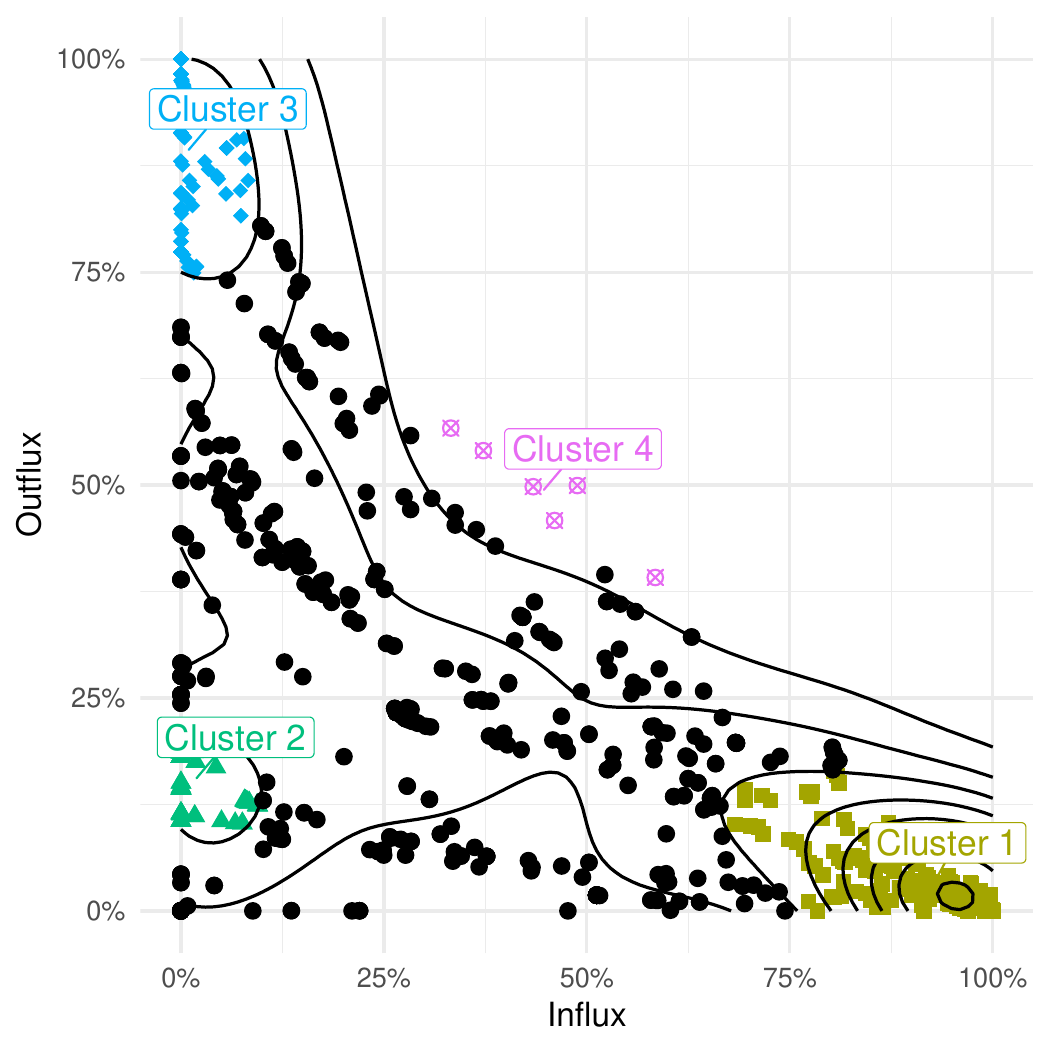}

}

\caption[Flux analysis plot.]{\label{fig-flux}Flux analysis plot: Influx
\(I(\mathbf{X}_{j})\) is shown on the x-axis and outflux
\(O(\mathbf{X}_{j})\) on the y-axis. Contours of a two-dimensional
density estimate are shown in the background.}

\end{figure}

These findings show that simple ad-hoc imputation strategies like mean
imputing e.g.~should not be applied. We recommend the proper separation
of the different types of observations (e. g. lab test types) before
imputation. The following results on a missing data structure analysis
can support researchers in defining a reasonable imputation strategy:

Good candidates for an imputation model are column pairs which have
non-missing values co-occurring with missing values or vice versa. These
two properties are measured with the outflux \(O(\mathbf{X}_{j})\) and
influx \(I(\mathbf{X}_{j})\). The results are shown in
Figure~\ref{fig-flux}.

Each column of each table is depicted as a point in
Figure~\ref{fig-flux}. The contours of a density estimate are included
to highlight clusters of columns. Based on a cluster analysis, we
identified four clusters of interest.

\begin{itemize}
\tightlist
\item
  Cluster 1: Missing values in columns of this cluster may be easy to be
  imputed by other columns, but the column itself may be not very useful
  to impute other columns.
\item
  Cluster 2: Missing values in columns of this cluster may by difficult
  to be imputed by other columns and the column itself maybe not very
  useful to impute other columns.
\item
  Cluster 3: Missing values in columns of this cluster may be difficult
  to be imputed by other columns, but the column itself my be useful to
  impute other columns.
\item
  Cluster 4 contains columns with both medium influx and outflux. These
  columns have many non-missing values co-occurring with missing values
  and vice versa. They might also be useful for imputation.
\end{itemize}

A list of the columns in clusters 3 and 4 is shown in Supplementary
Table 3, Additional File 1.

It is important to note here that influx and outflux only provide first
insights into the usability of a column for imputation without providing
information how efficient an imputation finally will be. However, a
pre-selection of columns based on the flux analysis can help to identify
columns which are more likely to be useful for imputation.

Missing values might be caused by known or unknown mechanisms in the
data collection and reporting process.

Table~\ref{tbl-inst-missing-rmse} shows columns which were important
predictors for the proportion of missing values per observation
\(OPM(\mathbf{X},\text{DP},i)\). The RMSE gives the prediction error on
the test set. Because analyses were performed separately for each
\(\text{DP}\), results are given for each \(\mathbf{X}\) and
\(\text{DP}\). Columns with an adjusted importance above 50\% were
considered an important predictor and are shown in this table. Full
results are available in Supplementary Table 4, Additional File 1.

\begin{longtable}[]{@{}
  >{\raggedright\arraybackslash}p{(\linewidth - 4\tabcolsep) * \real{0.4020}}
  >{\raggedleft\arraybackslash}p{(\linewidth - 4\tabcolsep) * \real{0.0980}}
  >{\raggedright\arraybackslash}p{(\linewidth - 4\tabcolsep) * \real{0.5000}}@{}}

\caption{\label{tbl-inst-missing-rmse}Important predictors for the
proportion of missing values in an observation in each table and
contributing data provider.}

\tabularnewline

\toprule\noalign{}
\begin{minipage}[b]{\linewidth}\raggedright
Table (DP)
\end{minipage} & \begin{minipage}[b]{\linewidth}\raggedleft
Test RMSE
\end{minipage} & \begin{minipage}[b]{\linewidth}\raggedright
Important Predictors
\end{minipage} \\
\midrule\noalign{}
\endhead
\bottomrule\noalign{}
\endlastfoot
\texttt{T\_Recip\_Immuno} (ET) & 0.029 & \texttt{{C\_ResultType,
C\_DiagCenter}} \\
\texttt{T\_Donor\_PM\_Diag} (DSO) & 0.033 & \texttt{{C\_DiagClass}} \\
\texttt{T\_Recip\_Urgency} (ET) & 0.055 & \texttt{{C\_UrgeCode}} \\
\texttt{T\_Donor\_PM\_Crossmatch} (DSO) & 0.062 &
\texttt{{C\_TXC, C\_XMERG}} \\
\texttt{T\_Donor\_PM\_Med} (DSO) & 0.071 & \texttt{{C\_DosUnit}} \\
\texttt{T\_Transplantation} (ET) & 0.089 & \texttt{{C\_FolUpCenter,
C\_RegCenter,
C\_TXCenter}} \\
\texttt{T\_Kidney\_Followup} (ET) & 0.102 & \texttt{{C\_FolUpCenter,
C\_FolUpType}} \\
\texttt{T\_Donor\_PM\_Blood\_Type} (DSO) & 0.115 &
\texttt{{C\_DiagType}} \\
\texttt{T\_Donor\_PM\_Monitor} (DSO) & 0.168 &
\texttt{{C\_ArrestDur}} \\

\end{longtable}

The decision tree analysis revealed patterns in the missing data
structure across all tables shown in Table~\ref{tbl-inst-missing-rmse}.
A clear finding emerged: columns identifying observation types (e.g.,
test type, diagnosis class, urgency code) and centers were able to
predict missing data patterns.

Type identifiers were the most important predictors in five of the nine
table-data-provider pairs. For example, in \texttt{T\_Donor\_PM\_Diag},
the diagnosis class \texttt{C\_DiagClass} (identifying one of ten
different diagnosis types) was the primary predictor. Similarly, in
\texttt{T\_Recip\_Immuno}, the lab result type \texttt{C\_ResultType}
was the strongest predictor, reflecting that different laboratory tests
use different sets of columns. In \texttt{T\_Recip\_Urgency}, the
urgency code \texttt{C\_UrgeCode} determined which additional
urgency-specific columns were completed. This highlights how important
it is to separate the different types of observations (e.g., lab test
types) before imputation or analysis.

Center identifiers were also prominent predictors, appearing in three
table-data-provider pairs. In \texttt{T\_Recip\_Immuno}, the diagnostic
center \texttt{C\_DiagCenter} was the second most important predictor
alongside the test type, indicating that different centers report
different sets of results. For \texttt{T\_Transplantation} and
\texttt{T\_Kidney\_Followup}, various center columns (follow-up center,
registration center, transplant center) predicted missingness,
suggesting center-specific reporting practices. Here, the centers might
perform different types of follow-ups or have varying protocols for data
collection. However, center identifiers in the exported dataset are
anonymized/pseudonymized (in accordance with §15g TPG), so linkage to
real transplant centers is not possible under the current data access
agreement \citep{txregHandbuch2025}. This should be considered when
interpreting center-level differences in missingness patterns.

Notable exceptions were the dosage unit in \texttt{T\_Donor\_PM\_Med}
and cardiac arrest duration in \texttt{T\_Donor\_PM\_Monitor}, which may
serve as surrogates for medication or monitoring type information. The
prediction varied, test RMSE values in the shown results ranged from
0.029 to 0.168, demonstrating that missing value patterns can be
reliably predicted using these structural variables.

\subsection{Data Consistency}\label{sec-results-cons}

\begin{figure}[H]

\centering{

\includegraphics[width=3.5in,height=3.5in]{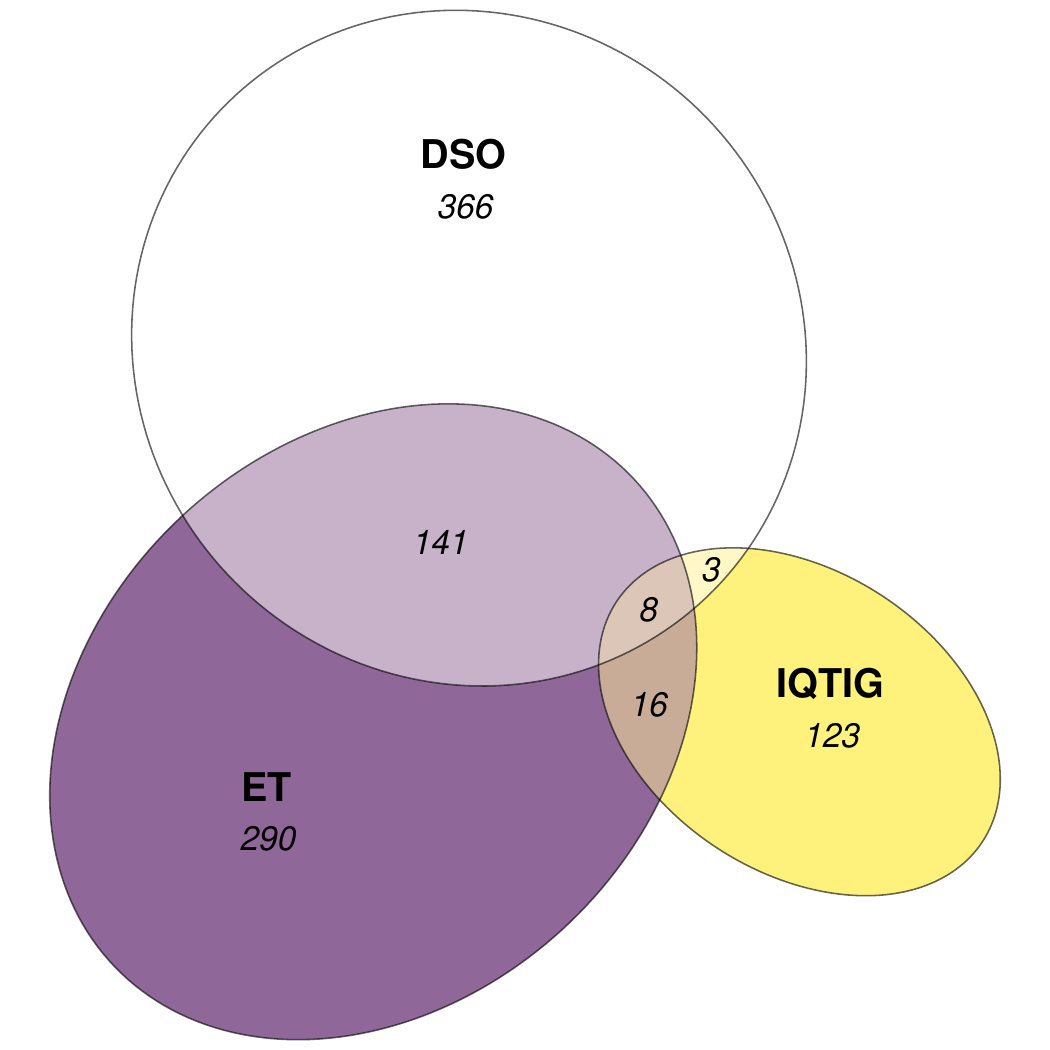}

}

\caption[Number of columns contributed by each data
provider.]{\label{fig-samename-cols}Number of columns contributed by
each data provider. Overlaps are multi-sourced variables.}

\end{figure}

First, we analyzed the multi-sourced variables.
Figure~\ref{fig-samename-cols} visualizes the number of columns that
represent these multi-sourced variables, grouped by the data provider.

Our analysis identified 168 multi-sourced variables across all tables.
Of those, 141 are provided by DSO and ET, while only 8 variables are
provided by all three data providers (DSO, ET, and IQTIG).

\begin{figure}[H]

\centering{

\includegraphics[width=3.5in,height=3.5in]{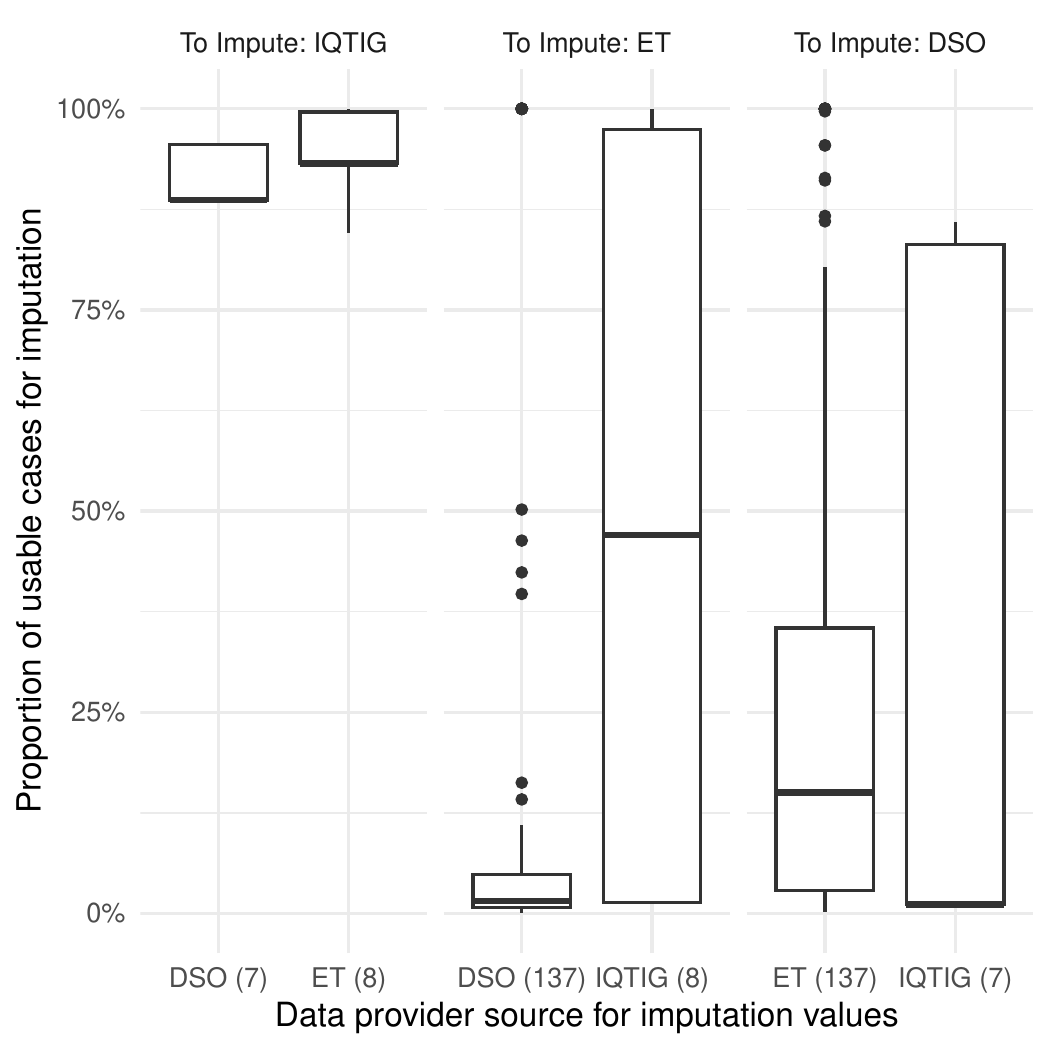}

}

\caption[Boplots of the usability of columns in multi-sourced variables
for imputation.]{\label{fig-red-missing}Boxplots of the usability of
columns in multi-sourced variables for imputation.}

\end{figure}

Next, we investigate the potential of the multi-sourced variables for
cross--data-provider imputation by analysing the proportion of usable
cases \(U(\mathbf{X}_j,\mathbf{X}_k)\). Figure~\ref{fig-red-missing}
displays the distribution of \(U(\mathbf{X}_j,\mathbf{X}_k)\) values
across pairs of multi-sourced variables. Only pairs are shown where both
data providers reported a value for at least one observation. In this
visualization, the y-axis represents \(U(\mathbf{X}_j,\mathbf{X}_k)\),
the x-axis shows the target data provider for imputation, and the panel
headers indicate the source data provider providing the imputation
values. As an example, for missing values in IQTIG columns, the median
proportion of usable cases for imputation when using corresponding DSO
columns as sources was 89\%, indicating high imputation potential.

The results reveal distinct patterns of imputation utility across
data-provider combinations. For missing values in multi-sourced IQTIG
columns, approximately 80\% of cases had corresponding non-missing
values in either DSO or ET columns, indicating these sources provide
substantial imputation opportunities for IQTIG data. Conversely,
imputation of multi-sourced ET columns using DSO sources showed limited
potential, with only few column pairs exhibiting proportions of usable
cases above 25\%. The observed outliers reaching 100\% usability
corresponded to entry dates and ET identifier columns.

IQTIG columns provided limited utility for imputing missing ET or DSO
values. The potential for imputting multi-sourced DSO columns using ET
sources is also limited (median 15\% usable cases) but still higher as
the reverse scenario of imputing ET columns from DSO sources (median 2\%
usable cases). Based on these findings, we recommend prioritizing ET and
DSO columns for primary analysis while utilizing IQTIG columns primarily
as sources for imputing missing values in ET and DSO datasets.

This imputation strategy is based on the assumption that all data
providers provide the same kind of information and thus give consistent
results. In the following we show inter--data-provider agreements for
all pairs of columns within multi-sourced variables. For 37 out of 184
pairs, no association could be calculated due to constant data, no
non-missing values occurring for the same observation or data type
mismatches.

Of the 147 column pairs for which we could calculate association
measures, 139 pairs showed a high agreement with a measured association
of 0.9 and above. The 8 pairs that exhibit associations below 0.9 and
thus some disagreement between data providers are shown in
Table~\ref{tbl-red-scores-outliers}. For example, the recipient's height
columns from the DSO and ET in \texttt{T\_Recipient} show an absolute
value of the Pearson correlation coefficient of only 0.01, that
indicates that DSO and ET systematically provide different information
for height.

\begin{longtable}[]{@{}
  >{\raggedright\arraybackslash}p{(\linewidth - 6\tabcolsep) * \real{0.3537}}
  >{\raggedright\arraybackslash}p{(\linewidth - 6\tabcolsep) * \real{0.3780}}
  >{\raggedright\arraybackslash}p{(\linewidth - 6\tabcolsep) * \real{0.1220}}
  >{\raggedleft\arraybackslash}p{(\linewidth - 6\tabcolsep) * \real{0.1463}}@{}}

\caption{\label{tbl-red-scores-outliers}Multi-sourced variables with
high degree of discrepancy between data providers.}

\tabularnewline

\toprule\noalign{}
\begin{minipage}[b]{\linewidth}\raggedright
Column
\end{minipage} & \begin{minipage}[b]{\linewidth}\raggedright
Table
\end{minipage} & \begin{minipage}[b]{\linewidth}\raggedright
DP Pair
\end{minipage} & \begin{minipage}[b]{\linewidth}\raggedleft
Association
\end{minipage} \\
\midrule\noalign{}
\endhead
\bottomrule\noalign{}
\endlastfoot
\texttt{C\_WarmIschemiaTime} & \texttt{T\_Kidney\_Retrieval} & DSO, ET &
0.009 \\
\texttt{C\_BodyHeight} & \texttt{T\_Recipient} & ET, IQTIG & 0.011 \\
\texttt{C\_BodyWeight} & \texttt{T\_Recipient} & ET, IQTIG & 0.054 \\
\texttt{C\_Hypertension} & \texttt{T\_Donor\_PM} & DSO, ET & 0.084 \\
\texttt{C\_DiuresisInterval} & \texttt{T\_Donor\_PM\_Monitor} & DSO, ET
& 0.154 \\
\texttt{C\_Malign} & \texttt{T\_Donor\_PM} & DSO, ET & 0.383 \\
\texttt{C\_Diabetes} & \texttt{T\_Donor\_PM} & DSO, ET & 0.388 \\
\texttt{C\_Creatinine} & \texttt{T\_Donor\_PM\_Chem} & DSO, ET &
0.561 \\

\end{longtable}

Results of these column pairs reveal specific patterns of discrepancies
between data providers. Body measurements shows particularly poor
agreement: height \texttt{T\_Recipient.C\_BodyHeight} and weight
\texttt{T\_Recipient.C\_BodyWeight} between ET and IQTIG suffer from
inconsistent unit usage and data errors in IQTIG data.

The warm ischemia time \texttt{T\_Kidney\_Retrieval.C\_WarmIschemiaTime}
shows minimal association (0.01) between DSO and ET, suggesting
different measurement protocols or timing definitions across data
providers. Similarly, diuresis duration
\texttt{T\_Donor\_PM\_Monitor.C\_DiuresisInterval} and creatinine values
\texttt{T\_Donor\_PM\_Chem.C\_Creatinine} demonstrate unit
inconsistencies, with DSO reporting apparently incorrect units leading
to physiologically implausible measurements. These inconsistencies are
plausibly attributable to historical differences in units used across
Germany (SI in Eastern Germany vs conventional in Western Germany),
which may create clerical errors during data entry
\citep{Kutschmann2013}.

The variables for hypertension \texttt{C\_Hypertension}, malignancy
\texttt{C\_Malign}, and diabetes mellitus \texttt{C\_Diabetes} in
\texttt{T\_Donor\_PM} showed low inter--data-provider associations due
to different categorical coding schemes employed by DSO and ET. Here in
addition, the DSO only documents present status for confirmed diseases,
absence is not explicitly recorded.

In summary, these findings show that for many multi-sourced variables,
pairwise imputations are feasible. However, we identified important
exceptions where manual inspections would be required before imputation.

\subsection{Distribution of Event Time
Data}\label{sec-results-eventtime}

\begin{figure}[H]

\centering{

\includegraphics[width=\linewidth,height=3.5in,keepaspectratio]{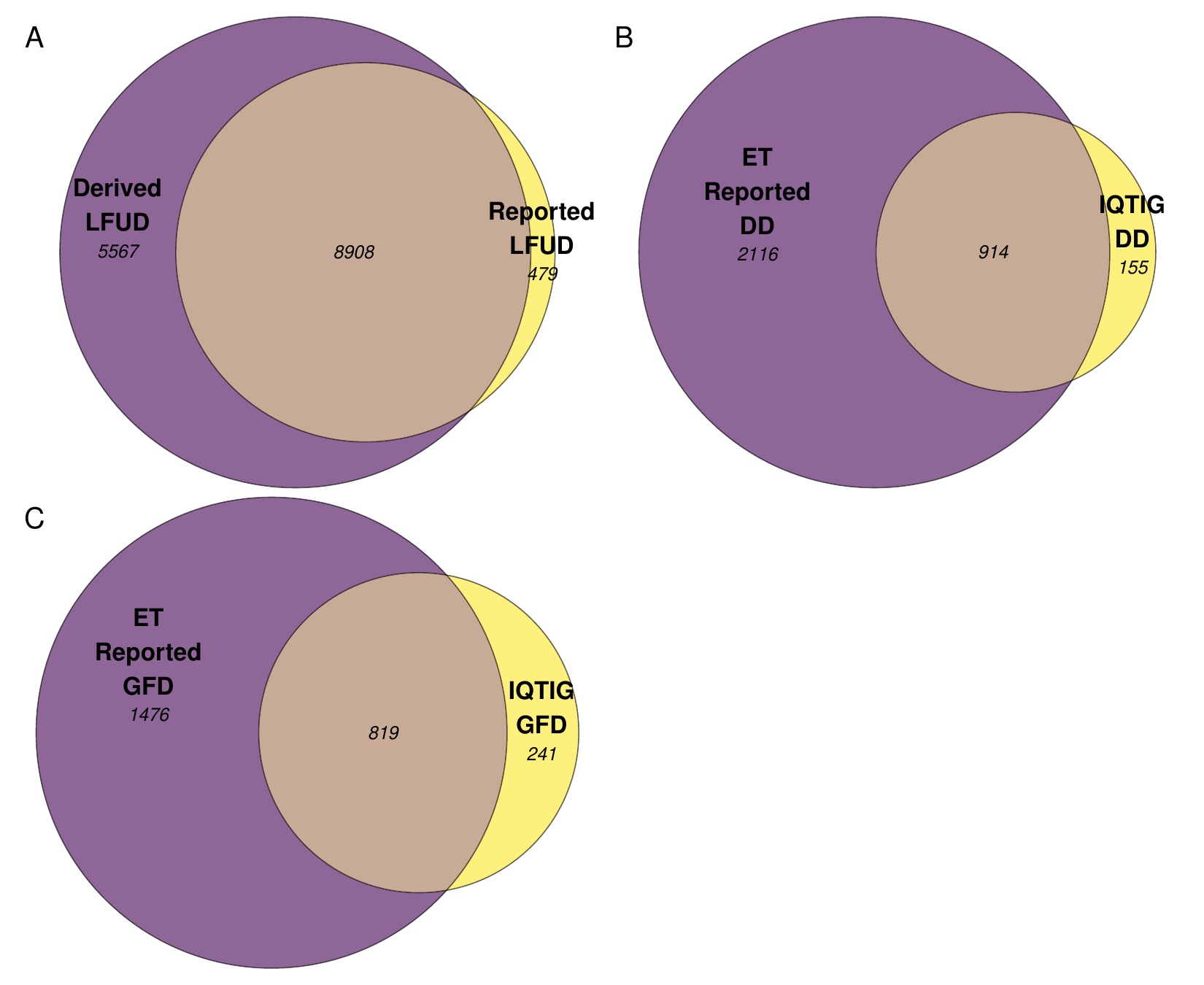}

}

\caption[Number of recipients for whom information on the date of last
follow-up (A), date of death (B) or date of graft failure (C) is
available, seperately for the different data providers that provide the
information.]{\label{fig-avail}Number of recipients for whom information
on the date of last follow-up (A), date of death (B) or date of graft
failure (C) is available, seperately for the different data providers
that provide the information.}

\end{figure}%

\subsubsection{Time to Last Follow-up}\label{time-to-last-follow-up}

There are multiple sources for the date of last follow-up in the
dataset. First, ET directly provides a date for the last follow-up in
\texttt{T\_Transplantation.C\_LastDate}. In the following, we will refer
to this date as the ``Reported Last Follow-Up Date'' (``Reported
LFUD''). Furthermore, ET and IQTIG each provide a series of
\(0,1,\ldots\) follow-up dates in multi-sourced variable columns
\texttt{T\_Kidney\_Followup.C\_Date}. The corresponding combined last
follow-up date can be defined as the latest date for each recipient. We
call this the ``Derived LFUD''. In contrast to the single ET source
``Reported LFUD'', this derived date is calculated from follow-up
records contributed by both ET and IQTIG in \texttt{T\_Kidney\_Followup}
and therefore reflects the latest available tracked follow-up entry in
that table. Consequently, ``Derived LFUD'' can still be earlier than
``Reported LFUD'' if ET reports a later follow-up date in
\texttt{T\_Transplantation.C\_LastDate} that is not present in follow-up
observations. Data provided by the DSO includes only donor data without
recipient follow-up. recipient follow-up.

In general, IQTIG collects follow-up data during the initial hospital
stay and one, two, and three years after transplantation, whereas ET may
provide additional data e.g.~collected in the context of a registration
for re-transplantation or -allocation. For both data providers,
information beyond year 3 after transplantation may therefore be
incomplete, but are included in the following to provide a full picture
of the data.

\begin{figure}[H]

\centering{

\includegraphics[width=\linewidth,height=3.5in,keepaspectratio]{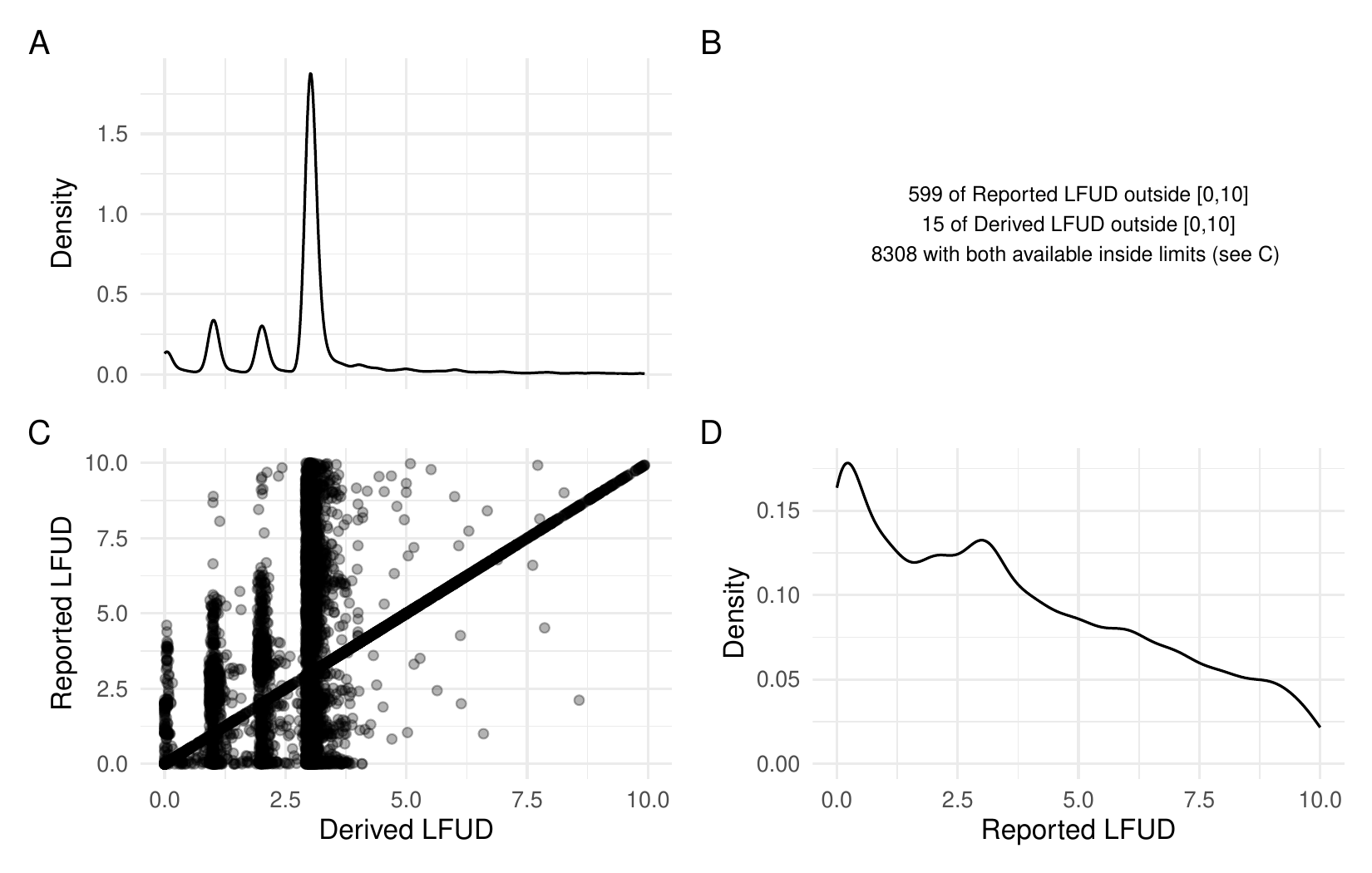}

}

\caption[Distribution of the follow-up dates reported by different data
providers.]{\label{fig-lfud-dates}Distribution of the follow-up dates
reported by different data providers.}

\end{figure}

Recipients may either have information on the ``Reported LFUD'', the
``Derived LFUD'' or both. This can be seen in Figure~\ref{fig-avail}.
96.8\% (14,475) of the recipients have a last follow-up date according
to the ``Derived LFUD''. The ``Reported LFUD'' is given for 62.8\%
(9,387) of the recipients including all 3\% (479) of the recipients with
no available ``Derived LFUD''. Figure~\ref{fig-lfud-dates} illustrates
the distribution of both dates (A and D) together with their association
(C).

In Figure~\ref{fig-lfud-dates} data of 15 and 599 observations are not
shown as these were lying before date of transplantation (defining
timepoint 0) or more than 10 years after transplantation.

The ``Derived LFUD'' is clustering at 1, 2, and 3 years (A). This can be
explained by the yearly follow-up schedule of patients after kidney
transplantation. The ``Reported LFUD'' has a more even distribution of
dates but also shows a peak at year 3 (D). As a consequence, the
association between both dates is low (C).

\subsubsection{Time to Patient Death}\label{sec-results-death}

There are two sources for a death date:

\begin{itemize}
\tightlist
\item
  ET provides the recipient death date in
  \texttt{T\_Recipient.C\_DeathDate}. We will refer to this date as the
  ``ET Reported Death Date'' (``ET Reported DD'').
\item
  In \texttt{T\_Kidney\_Followup.C\_DeathDate}, a death date is reported
  by IQTIG if the recipient has died since the last follow-up. In the
  case of multiple reported dates, the earliest will be used. We will
  refer to this as the ``IQTIG Reported Death Date'' (``IQTIG Reported
  DD'').
\end{itemize}

The same analysis as for the last follow-up dates was done for the death
dates. Figure~\ref{fig-avail} B shows that for 3,185 recipients at least
one of the three dates is given and most of them are only reported by ET
(2,116). The ``IQTIG Reported DD'' is observed for for 1,069 of all
recipients.

\begin{figure}[H]

\centering{

\includegraphics[width=\linewidth,height=3.5in,keepaspectratio]{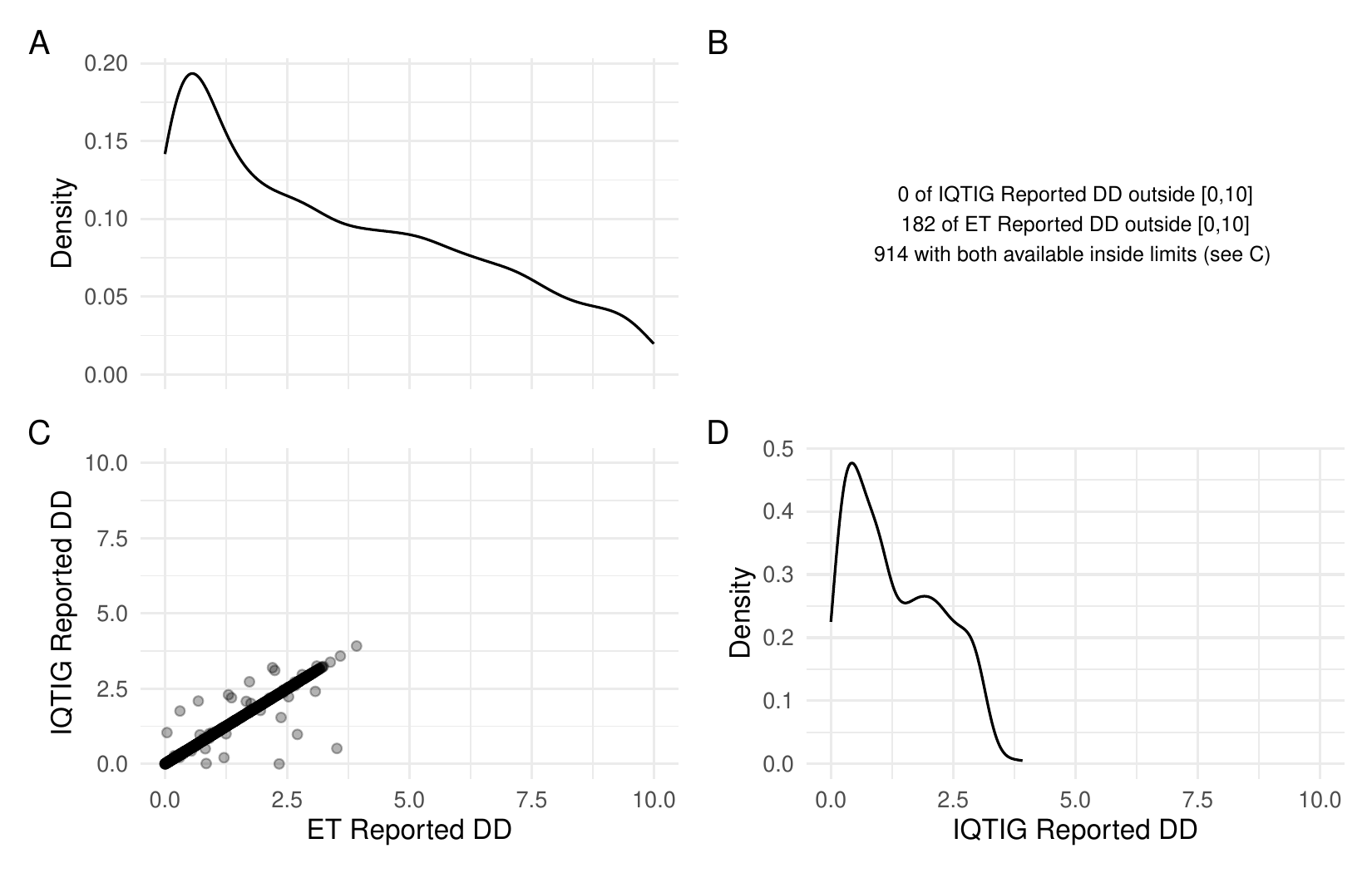}

}

\caption[Comparison of the death
dates.]{\label{fig-death-dates}Comparison of the death dates.}

\end{figure}

Figure~\ref{fig-death-dates} illustrates the distribution of the two
dates (A, D) together with their associations (C).

In Figure~\ref{fig-death-dates} data of 182 observations are not shown
as the ``ET Reported DD'' was reported lying before date of
transplantation (defining timepoint 0) or more than 10 years after
transplantation.

We observe that some death dates are reported beyond year 3, in
particular when reported by ET in ``ET Reported DD''. The date sources
``ET Reported DD'' and ``IQTIG Reported DD'' agree in most cases
(subfigure C).

\subsubsection{Time to Graft Failure}\label{sec-results-graftfailure}

Similar to recipient death, the date of graft failure is also provided
in two sources.

\begin{itemize}
\tightlist
\item
  ET provides the date of graft failure in
  \texttt{T\_Transplantation.C\_FailureDate} to which we refer as ``ET
  Reported Graft Failure Date'' (``ET Reported GFD'').
\item
  IQTIG provides a graft failure date in
  \texttt{T\_Kidney\_Followup.C\_FailureDate}, to which we refer as
  ``Derived Graft Failure Date'' (``Derived GFD'').
\end{itemize}

The analysis of the graft failure dates results in similar observations
as for the death dates. For 2,536 recipients at least one of the two
dates is given as can be seen in Figure~\ref{fig-avail} B. The ``IQTIG
Reported GFD'' is observed for 1,060 of all recipients. 1,476 of all
observed graft failures are only reported in ``ET Reported GFD'', while
241 are only reported by IQTIG.

\begin{figure}[H]

\centering{

\includegraphics[width=\linewidth,height=3.5in,keepaspectratio]{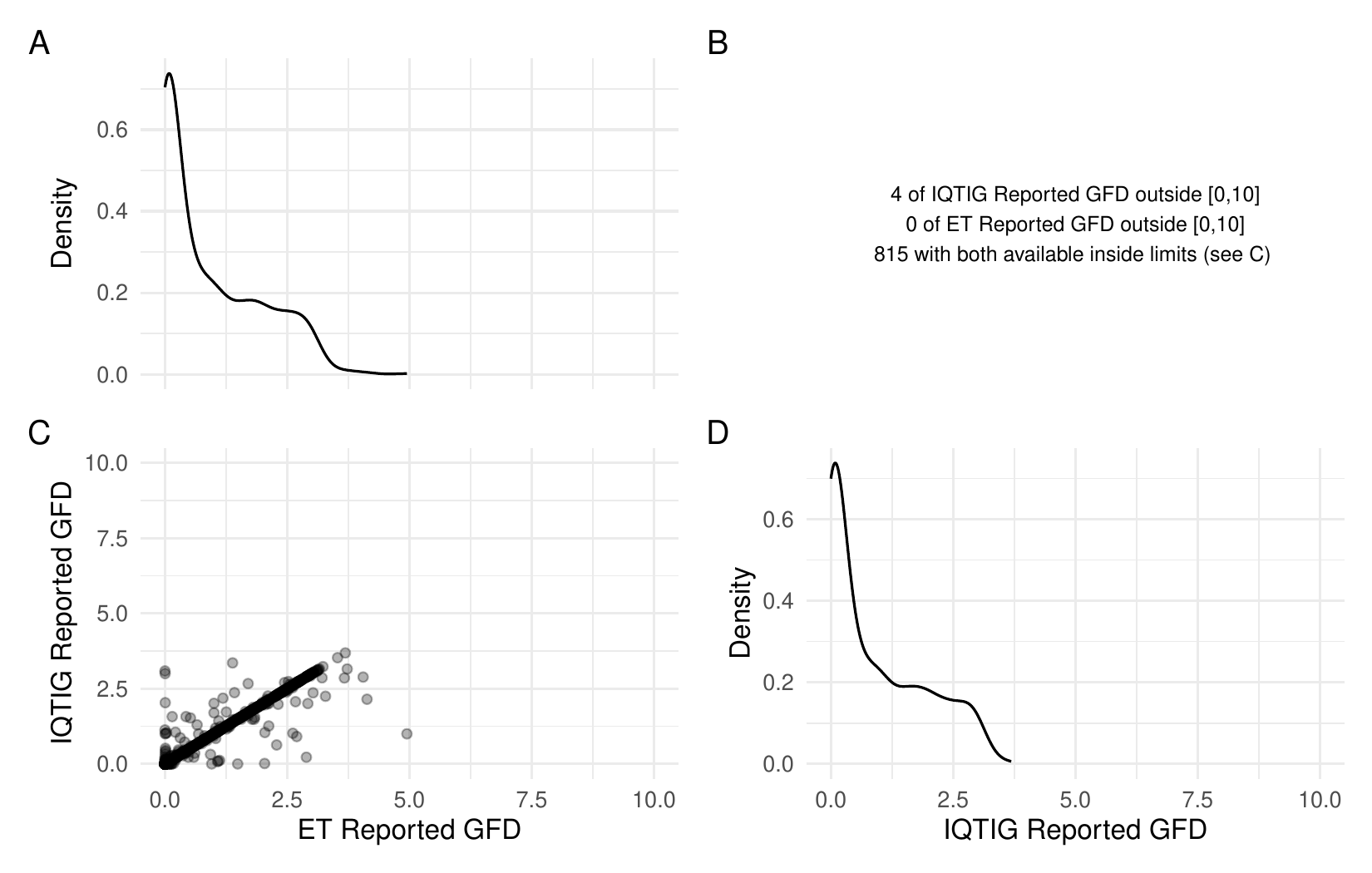}

}

\caption[Comparison of the graft failure
dates.]{\label{fig-gfd-dates}Comparison of the graft failure dates.}

\end{figure}

Figure~\ref{fig-gfd-dates} illustrates the distribution of the two dates
(A, D) together with their associations (C).

In Figure~\ref{fig-gfd-dates} data of 4 observations are not shown as
the ``ET Reported DD'' was repported lying before date of
transplantation (defining timepoint 0) or more than 10 years after
transplantation.

Results are similar to the findings for the dates of death as reported
from the different data providers:. While ``ET Reported GFD'' and
``IQTIG Reported GFD'' show some agreement in many cases there is a
higher number of cases, where the dates differ.

\subsubsection{Consequences of Different Outcome
Definitions}\label{consequences-of-different-outcome-definitions}

To illustrate the importance of carefully defining the event time
variables we estimate survival probabilities for the composite endpoint
graft and patient survival when constructing the outcome from different
sources.

We consider different combinations of the possible definitions for the
date of death (\(T^D\)), the date of graft failure (\(T^G\)) and the
date of last follow-up (\(T^F\)) as given in table
Table~\ref{tbl-outcome}.

\begin{longtable}[]{@{}
  >{\raggedright\arraybackslash}p{(\linewidth - 8\tabcolsep) * \real{0.1091}}
  >{\raggedright\arraybackslash}p{(\linewidth - 8\tabcolsep) * \real{0.2909}}
  >{\raggedright\arraybackslash}p{(\linewidth - 8\tabcolsep) * \real{0.3030}}
  >{\raggedright\arraybackslash}p{(\linewidth - 8\tabcolsep) * \real{0.2545}}
  >{\raggedright\arraybackslash}p{(\linewidth - 8\tabcolsep) * \real{0.0424}}@{}}

\caption{\label{tbl-outcome}Comparison of the different outcome
definitions.}

\tabularnewline

\toprule\noalign{}
\begin{minipage}[b]{\linewidth}\raggedright
Outcome
\end{minipage} & \begin{minipage}[b]{\linewidth}\raggedright
\(T^D\)
\end{minipage} & \begin{minipage}[b]{\linewidth}\raggedright
\(T^G\)
\end{minipage} & \begin{minipage}[b]{\linewidth}\raggedright
\(T^F\)
\end{minipage} & \begin{minipage}[b]{\linewidth}\raggedright
N
\end{minipage} \\
\midrule\noalign{}
\endhead
\bottomrule\noalign{}
\endlastfoot
a) ET Reported & ET Reported DD & ET Reported GFD & Reported LFUD &
10,937 \\
b) IQTIG Reported & IQTIG Reported DD & IQTIG Reported GFD & Derived
LFUD & 14,471 \\
c) Combined & Minimum of ET Reported DD and IQTIG Reported DD & Minimum
of ET Reported GFD and IQTIG Reported GFD & Maximum of Reported LFUD and
Derived LFUD & 14,950 \\

\end{longtable}

For each recipient \(i\) we define the time to composite outcome or
censoring, whatever comes first as

\[
T_i = \min(T^{D}_{i},T^{G}_{i},T^{F}_{i},3),
\]

separately for the different combinations of possible definitions of
dates as given in Table~\ref{tbl-outcome}. If \(T^D_i\), \(T^G_i\), and
\(T^F_i\) are missing, \(T_i\) will be defined as missing and excluded
from analyses. Otherwise the minimum is derived from the non-missing
information only. The variable \(\Delta_i\) indicates if \(T_i\) refers
to an event or a censored observation and is defined as

\[
\Delta_i = \mathds{1}\left(T_i = \min(T^D_i,T^G_i)\right).
\]

By only considering dates of death and graft failure within the first 3
years after transplantation, we consider the findings indicating that
information beyond year 3 may be unreliable.

For the analysis, implausible event or follow-up dates were handled
using predefined plausibility rules. Recipients with at least one event
or follow-up date reported more than 30 days (\textgreater1 month)
before the transplantation date were excluded (4). Reported dates
occurring within 30 days prior to transplantation were recoded to the
transplantation date (affecting 6 dates). Dates occurring 15 years or
more after transplantation were set to missing (affecting 5 dates in
``Reported LFUD'').

Table~\ref{tbl-outcome} also gives the number of recipients for each
combination. In all but one case, data are available for around 14,500
recipients. The only exception is ``a) ET Reported'' where both event
and censoring dates are missing in around 4,000 cases.

\begin{figure}[H]

\centering{

\includegraphics[width=\linewidth,height=3.5in,keepaspectratio]{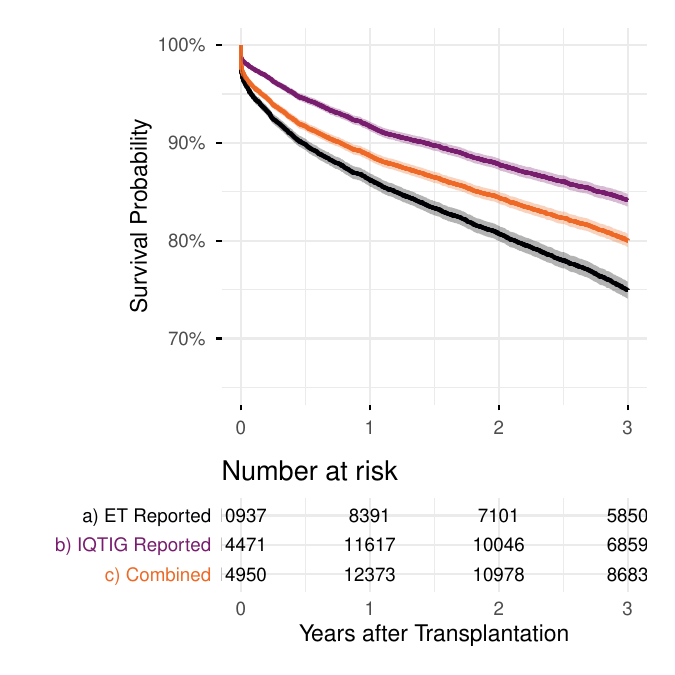}

}

\caption[Kaplan-Meier estimates of graft and patient survival
probability over time.]{\label{fig-survival-all}Kaplan-Meier estimates
of graft and patient survival probability over time (composite
endpoint).}

\end{figure}

Kaplan-Meier estimates \(\hat{S}\) for the composite endpoint of graft
and patient survival are calculated separately for each of the three
outcome definitions. Results are shown in Figure~\ref{fig-survival-all}.

\begin{itemize}
\tightlist
\item
  When relying on information provided by ET only (``a) ET Reported''),
  smaller survival probabilities are estimated than with all other
  outcomes. This can be explained by the fact that ET reports the
  largest number of patient deaths and graft failures, while about 4,000
  recipients without follow-up information are excluded. These likely
  refer mainly to survivors.
\item
  Conversely, when relying mainly on information provided by IQTIG (``b)
  IQTIG Reported''), higher survival probabilities are estimated
  compared to the results using information from ET. This can be
  explained by the fact that many patient deaths and graft failures are
  not reported by IQTIG.
\item
  The outcome ``c) Combined'' uses all information irrespective of the
  data provider that provided that information. As a consequence, the
  survival estimates lie between the estimates that rely on a single
  data provider only.
\end{itemize}

In summary, we therefore recommend using the ``c) Combined'' outcome
definition for event time analyses of single or composite events.

\section{Discussion}\label{sec-discussion}

Based on the finding of our initial data analysis, we propose the
following recommendations for data preprocessing and analysis:

\begin{enumerate}
\def\labelenumi{\arabic{enumi}.}
\tightlist
\item
  Tables containing a mix of cross-sectional and longitudinal data
  should be split by observation type prior to any imputation.
\item
  Flux analysis can be used to preselect variables for imputation,
  thereby helping to prioritize and focus efforts.
\item
  Multi-sourced variables should be used for imputing missing data. For
  important exceptions as indicated in this paper, we recommend to first
  harmonize units and codings before imputation using plausible value
  ranges and mappings.
\item
  Data from Eurotransplant (ET) and the German Organ Transplantation
  Foundation (DSO) should be prioritized for primary analyses, while
  data from the Institute for Quality Assurance and Transparency in
  Health Care (IQTIG) may serve as a supplementary source for
  imputation. This is in line with the role of IQTIG as a quality
  assurance institution.
\item
  When defining outcomes for time-to-event analysis, we recommend using
  a combined outcome that integrates information from all three data
  providers. Relying solely on data from a single source may lead to
  biased results.
\end{enumerate}

Other studies using the TxReg data have employed different strategies
for extracting the relevant information.
\citet{vonSamsonHimmelstjerna2023} analyzed patient time on the waiting
list for a first kidney transplantation in the ``legacy data'' between
2006 and 2016. While they reported using data from ET and DSO, they did
not mention the use of IQTIG data. Their study population included
24,150 recipients, which is larger compared to our population of 14,954
recipients. This is probably due to different inclusion criteria.
\citet{Kolbrink2024} combined both ``legacy data'' and ``new data'' to
analyze the impact of the 65-year age criterion for the ``ET Senior
Program'' (ESP) versus the standard ET Kidney Allocation System (ETKAS)
on time spent on the kidney transplant waiting list. They also did not
reference the use of IQTIG data. \citet{Otto2024} focused on data
quality in the ``legacy data'', also highlighting challenges related to
multi-sourced variables and emphasizing that the different data
providers collect data at different points in the transplantation and
follow-up process. Their analysis, which covered all organ types, noted
that not all recipients, donors, and transplantations are represented
consistently across all tables. For our study, detailed methodology
documentation for each source database (ET, DSO, IQTIG) was only
partially available. We therefore relied on the TxReg handbook and
publicly available registry documentation \citep{txregHandbuch2025}. The
limited transparency of source-specific documentation is itself an
important limitation of the registry and is consistent with the
challenges reported by \citet{Otto2024}.

In our study, we addressed this issue by limiting the target population
to recipients with consistently documented transplantations, donors, and
follow-ups across all relevant sources (see Supplementary Figures 2 to
4, Additional File 1). Before IQTIG was responsible for collecting the
quality assurance data, the BQS Institute had similar role. In studies
using DSO data from 2006 and 2008 - linked to BQS data - challenges were
reported similar to those we observed in the registry
\citep{Kutschmann2013, FischerFrhlich2015, Frhauf2011}.

\citet{Coemans2018} reported a one-year death-censored graft survival of
92.0\% for patients transplanted between 2006 and 2015. Using our
combined outcome variable (``c) Combined''), we obtained an estimated
one-year survival of 88.68\% for one year and a three-year survival of
80.03\%. For Iranian kidney transplant recipients,
\citet{GhelichiGhojogh2021} reported one year survival estimates of
92.48\% for graft survival and 91.27\% for patient survival. For three
year survival, the reported estimates were 85.08\% and 86.46\%.

Mortality ascertainment data in the TxReg is based on routine
documentation at transplant centers (and/or aftercare providers) and
subsequent reporting to the respective data providers (ET and IQTIG)
\citep{txregHandbuch2025}. This is different from for example the SRTR
database in the US, which is supplemented by the Social Security Death
Master File (SSDMF) \citep{srtrDatabase}. The usage of relative dates in
the TxReg instead of absolute dates, limits the potential analysis to
external events like the COVID-19 pandemic. this also prevents robust
calendar-time stratification into transplantation eras (e.g., 5-year
cohorts). As a result, together with the short follow-up, analyses
cannot be directly aligned to contemporary practice periods, which
limits immediate clinical interpretability.

For clinicians and patients, the central question is what the TxReg can
and cannot reliably support in practice.

Strengths of the TxReg data include: (1) broad national coverage across
key steps of kidney transplantation care, (2) complementary information
from ET, DSO, and IQTIG that can improve endpoint ascertainment when
sources are combined, and (3) the possibility to reconstruct recipient-,
donor-, and transplantation-level trajectories when records are
consistently linked.

Important limitations include: (1) insufficient documentation of data
collection processes and individual variable definitions across
providers, which hampers transparent interpretation, (2) inconsistent
representation of recipients, donors, and transplantations across and
within tables, requiring strict inclusion criteria to ensure comparable
data, and (3) missing reliable long-term follow-up data.

Accordingly, this IDA supports three robust practical conclusions:
first, TxReg data can be used for clinically relevant kidney-transplant
analyses when cohorts are restricted to consistently linked recipient,
donor, and transplantation records; second, endpoint ascertainment is
more complete when information from ET, DSO, and IQTIG is combined
rather than relying on a single source; third, reproducible inference
requires explicit preprocessing decisions, including source
prioritization, harmonized coding/units, and transparent handling of
missingness before model-based analyses.

Beyond these registry-level constraints, our study has additional
limitations specific to its scope: we focused on kidney transplantation
in the ``legacy data'' and did not include other organs or the
structurally different ``new data'' (post-2016). Some challenges,
including approximate record linkage with repeat transplantations, were
outside the present scope. Kaplan-Meier analyses were used
illustratively; for formal time-to-event inference in this setting,
competing-risk methods are more appropriate, as emphasized by
\citet{Coemans2025}.

\section{Conclusions}\label{sec-conclusion}

We demonstrated that using the TxReg data for research requires careful
consideration for data-pre-processing, handling missing values and
defining long-term outcomes. The reason is a complex data structure
caused by different data providers providing both distinct as well as
overlapping information. We identified 168 multi-sourced variables
reported by multiple providers in parallel, which introduce
discrepancies in coding and units but also provide opportunities for
missing data imputation. Furthermore, missing data patterns are often
predictable from observation types which can guide imputation strategies
in future analyses.

Our initial data analysis will assist researchers in both understanding
and making appropriate use of the TxReg data. In practical terms, our
analyses support three actionable conclusions: (1) kidney-transplant
analyses are feasible with consistently linked recipient, donor, and
transplantation records, but require careful preprocessing to account
for missing data and data inconsistencies, (2) combined ET/DSO/IQTIG
definitions improve ascertainment compared to single-source definitions,
and (3) future analyses using TxReg data should be preceded by a
targeted initial data analysis to ensure appropriate handling of the
registry's structural complexities.

\section{List of abbreviations}\label{list-of-abbreviations}

\begin{itemize}
\tightlist
\item
  TxReg: German Transplantation Registry
\item
  ET: Eurotransplant Foundation
\item
  DSO: German Organ Transplantation Foundation
\item
  IQTIG: Institute for Quality Assurance and Transparency in Health Care
\item
  IDA: Initial Data Analysis
\item
  ESKD: End-Stage Kidney Disease
\item
  CKD: Chronic Kidney Disease
\item
  DP: Data Provider
\end{itemize}

\section{Declarations}\label{declarations}

\subsection{Ethics approval and consent to
participate}\label{ethics-approval-and-consent-to-participate}

This study received ethical approval from the Ethics Committee of the
medical school of the Martin-Luther- University Halle-Wittenberg
(approval 2024-139) on August 06, 2024. This is an IRB-approved
retrospective study, all patient information was de-identified and
patient consent was not required. Patient data will not be shared with
third parties.

\subsection{Consent for publication}\label{consent-for-publication}

Not applicable.

\subsection{Availability of data and
materials}\label{availability-of-data-and-materials}

The data analysis presented in this manuscript uses retrospective and
anonymized data of the German Transplant Registry. These data have been
supplied by the Transplant Registry Agency represented by the
Gesundheitsforen Leipzig GmbH on the 11th of February 2022. Data can be
requested for research purposes here:
https://transplantations-register.de/en/research. The datasets generated
and/or analyzed during the current study are not publicly available due
to data protection regulations. Digital Supplementary materials are
available at https://github.com/Finesim97/txreg\_ida\_digital.

\subsection{Competing interests}\label{competing-interests}

The authors declared no potential conflicts of interest with respect to
the research, authorship, and/or publication of this article.

\subsection{Funding}\label{funding}

The research project was funded by the Federal Ministry of Education and
Research (project 13FH019KX1) and the German federal state of Hesse.

The publication was funded by Darmstadt University of Applied Sciences.

\subsection{Authors' contributions}\label{authors-contributions}

LK, GG, HS and AJ designed the study. LK performed the data analysis and
drafted the manuscript. HS contributed to the data analysis and revised
the manuscript. AJ, GG and AW supervised the project and revised the
manuscript. AR, CLFF and AW contributed to the interpretation of the
results and revised the manuscript. All authors read and approved the
final manuscript.

\subsection{Acknowledgements}\label{acknowledgements}

Not applicable.

\section{Additional files}\label{additional-files}

\begin{itemize}
\tightlist
\item
  \texttt{Additional\ File\ 1.pdf}:

  \begin{itemize}
  \tightlist
  \item
    File format: PDF
  \item
    Title: Supplementary Material
  \item
    Description: This file contains supplementary figures and tables
    referenced in the main text.
  \end{itemize}
\end{itemize}

\renewcommand\refname{References}
\bibliography{bibliography.bib}

\end{document}